\documentclass{article}
\usepackage{cite}
\usepackage{amsmath,amssymb,amsfonts}
\usepackage{algorithmic}
\usepackage[dvipsnames]{xcolor}
\usepackage{graphicx}
\usepackage{dblfloatfix}
\usepackage{textcomp}
\usepackage{wrapfig}
\usepackage{caption}
\usepackage{color,soul}
\usepackage{siunitx}
\usepackage{url}
\usepackage[caption=false,font=footnotesize]{subfig}
\usepackage{listings}
\usepackage{datetime2}
\usepackage{float}
\usepackage{comment}
\usepackage{multirow}
\usepackage{ulem}
\usepackage{todonotes}
\usepackage{mathtools}
\usepackage{changebar}

\def\BibTeX{{\rm B\kern-.05em{\sc i\kern-.025em b}\kern-.08em
    T\kern-.1667em\lower.7ex\hbox{E}\kern-.125emX}}
\definecolor{abstractbg}{rgb}{0.89804,0.94510,0.83137}
\date{}
\begin{document}
\title{A Personalised Formal Verification Framework for Monitoring Activities of Daily Living of Older Adults Living Independently in Their Homes}
\author{Ricardo Contreras, Filip Smola, Nuša Farič, Jiawei Zheng, \\Jane Hillston, and Jacques D. Fleuriot
\thanks{This research was funded by the Legal \& General Group (research grant to establish the independent Advanced Care Research Centre at the University of Edinburgh). The funder had no role in conducting the study, interpretation or the decision to submit for publication. The views expressed are those of the authors and not necessarily those of Legal \& General.}
\thanks{Ricardo Contreras, Filip Smola, Nuša Farič, Jane Hillston, and Jacques D. Fleuriot,  are with the School of Informatics,
The University of Edinburgh, 10 Crichton Street, Edinburgh, EH8 9AB, UK
(e-mail: rcontre2@ed.ac.uk; f.smola@ed.ac.uk; nfaric@ed.ac.uk;
jane.hillston@ed.ac.uk; jdf@ed.ac.uk). Jiawei Zheng is with DigitLab, University of Exeter (e-mail: j.zheng2@exeter.ac.uk) and worked on the project while at the University of Edinburgh.}
}

\maketitle

\begin{abstract}
There is an urgent need to provide quality-of-life to a growing population of older adults living independently.  Solutions that focus on the person and take into account their preferences and context are recognised as key.
We introduce a framework for representing and reasoning about the Activities of Daily Living of older adults living independently at home. The framework integrates data from sensors and data from participants derived from semi-structured interviews, home layouts and additional contextual information, such as the researchers' observations. These data are used to create formal models, personalised for each participant according to their preferences and context. Requirements specific to each individual are formulated and encoded in Linear Temporal Logic, and a model checker is used to verify whether each is satisfied by the model of the participant's behaviour. We demonstrate the framework’s generalisability by applying it to two different participants, highlighting its potential to enhance the safety and well-being of older adults ageing in place.
\end{abstract}

\section{Introduction}
\label{sec:introduction}
Advancements in medical science have significantly increased life expectancy in recent decades, resulting in a growing population of older adults (OAs)~\cite{Chen2023}. As an increasing number of such adults live independently~\cite{Tkatch2020ReducingLA}, greater attention is being paid to ensuring their safety, addressing their care needs and maintaining their quality-of-life (QoL)~\cite{Lette2020}. Innovative technologies are increasingly needed to support independent living and enhance the QoL of OAs.

Activities of Daily Living (ADLs) are essential self-care physical tasks that an individual performs on a regular basis. ADLs are fundamental for maintaining health and well-being \cite{Khalili2024}. Unexpected behaviour during these activities may indicate a decline in cognitive or physical abilities, potentially leading to unsafe situations~\cite{mmilnac2016}. Moreover, monitoring such activities could be crucial for detecting changes that may impact an individual’s QoL\@.  Sensors provide a powerful means of monitoring ADLs and, by extension, detecting deviations, thus potentially facilitating timely interventions that can safeguard or even improve the QoL of OAs.

In this paper, a novel framework is presented that combines data from sensors, interviews, home layout and sociological observations (i.e.\ the systematic analysis of social behaviours and interactions) for representing and reasoning about ADLs in the homes of OAs living independently.  The framework leverages unobtrusive, commercially-available sensors to capture ADL events (e.g.\ opening a refrigerator, closing a shower cubicle door).  These events are integrated through an automation platform and stored in a time-series database. To enrich the sensor data, contextual information about daily routines is systematically collected through interviews and sociological observations. This contextual data also enables the customisation of sensor setups and the creation of accurate, person-centred models that account for home layout, privacy preferences, and individual requirements.

This work is specifically aimed at OAs living alone in their homes, with the overall goal of identifying unanticipated behaviours over time.

By modelling the various actions involved in carrying out ADLs as a finite state model~\cite{clarke1999model}, we can form an abstraction of the behaviour of the OA\@. This enables the application of formal methods to verify whether the observed behaviour, captured by a set of sensors, satisfies a set of behavioural properties specific to that OA\@. These properties are derived from existing general guidelines and from specific requirements of the OA\@. Model checking \cite{clarke1999model}, a formal methods technique, is used to verify whether the properties hold for the observed behaviour. More specifically, the model checker NuSMV~\cite{cimattiNuSMV} is used to specify and verify the model, encoding properties in a dialect of Linear Temporal Logic (LTL) \cite{manna1992}.

This study builds upon our previous work~\cite{contreras2024} which involved using public datasets~\cite{datasetMunguia2004}. It significantly extends it through the use of our own data, collected from multiple participants within an OA population in Edinburgh, Scotland. Using the combined sensor and contextual data, we automatically generate models based on the placement of sensors in each participant's home, while properties are manually encoded to reflect the unique requirements of each participant. 

The main contributions of this work are as follows:

\begin{enumerate}
    \item A comprehensive framework integrating formal model verification with commercial sensors to capture ADL data in the homes of OAs.  This is then extended with explicit contextual data.  This facilitates the creation of personalised models and property specifications, linking ADL sensor events to OA preferences and needs.
    \item A rigorous and systematic approach for the specification of models and properties based on information acquired from semi-structured interviews with each participant
    \item Our approach is generalisable and has been applied to a diverse range of OAs.  Here we illustrate it with two participants with contrasting ADLs.
\end{enumerate}

This paper is organised as follows: Section~\ref{sec:RelatedWork} discusses related work in sensor-based ADL monitoring and formal methods. Section~\ref{sec:SystemCompDataCollection} describes the main components of the framework and data collection. Section~\ref{sec:DataProcessAndModels} describes the data processing and personalised models. Section~\ref{sec:ExperimentAnalysisResults} describes the experiments and the results. In Section~\ref{sec:Conclusion} we present our conclusions and outline future work.

\section{Related Work}
\label{sec:RelatedWork}
Early work on ADL monitoring using unobtrusive sensors has primarily focused on recognising and tracking ADLs. Initial studies, such as those by Munguia et al.~\cite{munguiatapia2003,munguiatapia2004}, used contact sensors and energy switches to capture events involving people living alone with the aim of tracking daily activities. This was expanded upon by teams deploying sensors in the homes of OAs to monitor ADLs specifically for health care purposes. Rantz et al.~\cite{Rantz2013-ra} used temperature, motion, pulse and respiration sensors to generate alerts notifying health care providers of potential illness or functional decline. Kaye et al.~\cite{Kaye2011-ft,Kaye2017-gg} used sensor systems in the homes of OAs to capture health related metrics as well as other measures of interest, such as mobility and sleep behaviour, in real-time. Their subsequent work~\cite{Kaye2018-rd} outlined a protocol for sensor deployment in homes focused primarily on technical aspects of installation and system functionality. While such technical rigour is essential, optimal sensor placement also requires user-centred approaches that consider individual routines and preferences. 
Skubic et al.~\cite{Skubic2015-di} developed a continuous in-home monitoring system for OAs using unobtrusive motion and bed sensors. Their system generates alerts when statistically significant changes indicate a potential illness or functional decline.  More recently, Janes et al.~\cite{Janes2025-zw} discussed the use of in-home sensor systems combined with health records to slow functional decline in people living with lateral sclerosis. Similarly, Sprint et al.~\cite{cooksMCI} explored how unobtrusive home sensors can be leveraged to monitor daily activities and assess cognitive impairments. 

While the aforementioned work, frequently conducted with clinical support, has primarily focused on health care, our work complements this perspective by addressing the quality of life of OAs by verifying that their routines are performed in accordance with safety and behavioural properties in combination with their own preferences.
Building on this distinction, subsequent studies have explored various aspects of ADL monitoring beyond health care contexts. For instance, Kang et al.~\cite{kang2023} focused on the use of IoT devices and sensor fusion to capture ADL events and improve data reliability, while Matsui et al.~\cite{matsui2020} proposed semi-automated mechanisms that relied on physical buttons for activity annotation. Tonkin et al.~\cite{Tonkin2023} worked on a multi-sensor dataset collected in a controlled residential setting, enhancing its accessibility through improved documentation, annotations, and organisation. That dataset comprises short-duration scripted ADLs. 

The use of formal methods in ADL monitoring has also been explored in previous studies. Konios et al.~\cite{konios2018} used Petri nets to distinguish between normal and abnormal behaviour in data captured from motion, light, temperature, and contact sensors. Garcia-Constantino et al.~\cite{garcia-Constantino2020,garciaPetri} monitored ambient and wearable sensor data to identify deviations from normal behaviour during preparation of kitchen beverages, also using Petri nets. Magherini et al.~\cite{magherini2013} used temporal logic to automate the recognition of three ADLs within a hypothetical scenario involving a person living alone. The activities assumed specific sequences and timing of actions, which were captured by camera sensors. A shared characteristic of the above approaches is their reliance on common assumptions about the ADLs a person performs and where these activities are conducted.  Moreover, these studies, involving formal verification and ADLs, were carried out in controlled, simulated environments. 

As discussed by Kessler et al.~\cite{Kessler2025-gs} and Alexander et al.~\cite{Alexander2025-io}, a person-centred approach is increasingly recognised as important when recommending and using technology to monitor OAs’ daily activities. Other examples of person-centred tailoring in different domains include work on activity recognition by Guerra et al.~\cite{guerra2023} and on supporting better self-management of health information for OAs by Robinson et al.~\cite{robinson2020}. In our work, we focus on OAs performing ADLs in their own homes and personalise the approach so that the data collection, models and properties to be verified are all tailored to their situation and preferences.

\section{Framework Components and Data Collection}
\label{sec:SystemCompDataCollection}

This section provides an overview of the main components used in our framework and the data collection process. 

\begin{figure*}[htbp]
  \centering
  \includegraphics[width=\textwidth]{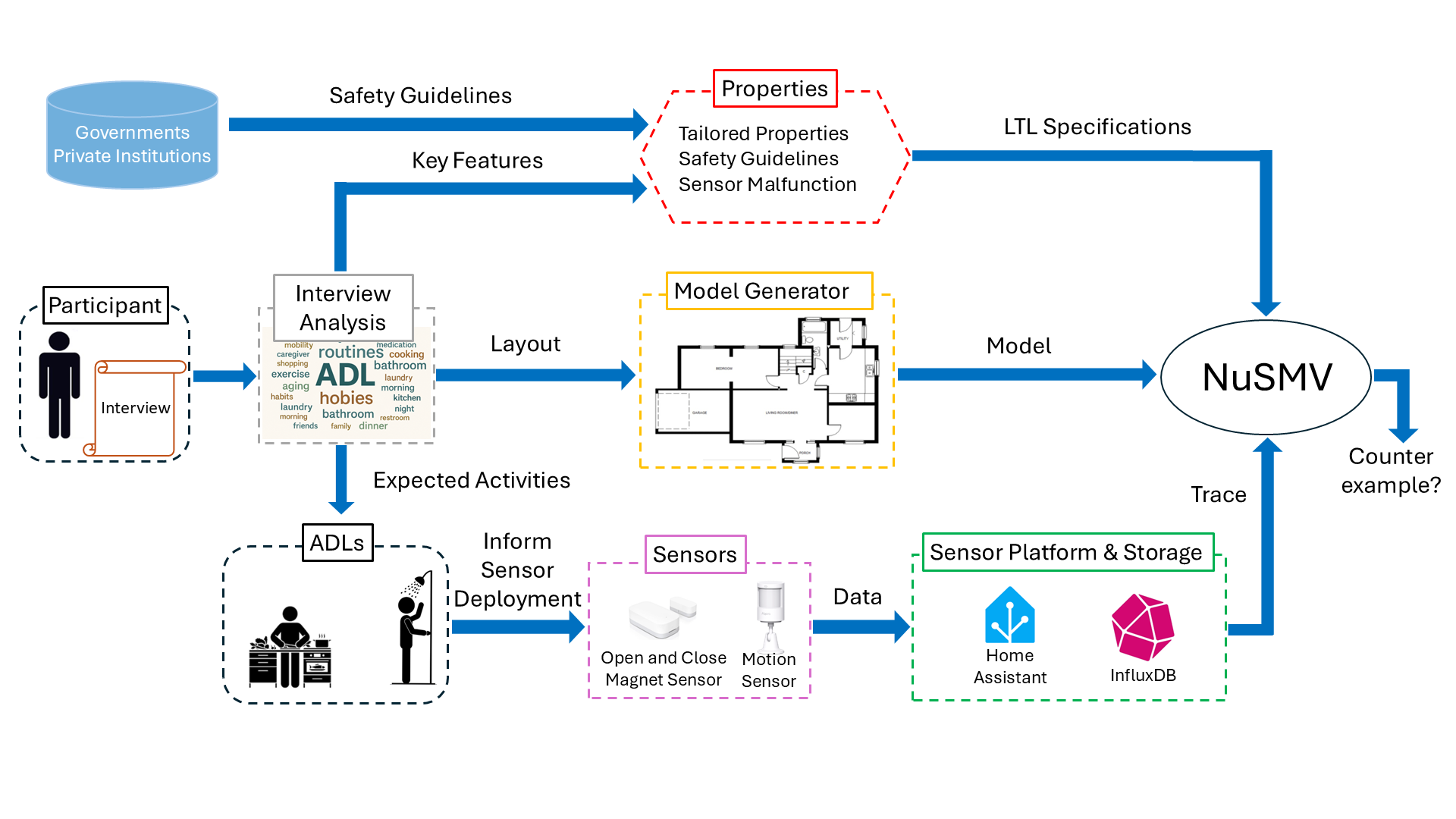}
  \caption{Main components of the framework.
  }
  \label{fig:achitecture}
\end{figure*}

\subsection{Framework Overview}

The main aspects of our framework are illustrated in Figure~\ref{fig:achitecture}. They are categorised into two groups of elements: 1) those involved during data collection in the participant's home and 2) those used after data collection.
The components present during data collection are:
\begin{itemize}
    \item Participants: Our study is a person-centric approach. Through semi-structured interviews, participants provide insights into their daily routines, preferences and specific needs, which will later be expressed as LTL properties. This also allows us to personalise the placement of sensors, by considering individual needs, their living environment and other aspects that may otherwise be overlooked (see Section~\ref{sub:DataCollectionProcess}).
    \item  Sensors: Two types of sensors are deployed in each participant’s home: contact and motion. Contact sensors monitor states related to doors, windows and appliances, providing data on object interactions. Motion sensors detect movement or changes in the position of participants within specific areas of the home. All sensors operate over the Zigbee wireless communication protocol~\cite{zigbee2020}.
    \item Sensor Platform and Storage: We employ open-source solutions for sensor data acquisition and management. Home Assistant~\cite{homeassistant}, a home automation platform, orchestrates data acquisition from our sensors. This provides a flexible and extensible setup for data aggregation. The sensor-generated data are stored in InfluxDB~\cite{influxdb}, a high-performance time-series database optimised for handling large volumes of time-stamped data. 
\end{itemize}
The components utilised post data collection are:

\begin{itemize}
    \item Interview Analysis: Interviews are parsed to identify participants’ routines and preferences. Qualitative analysis informs property specification, highlighting relevant aspects of personalisation, such as environment and needs (see Section~\ref{subsec:Preprocessing}).
    \item Model Generator: This generates a formal model of the participant's home environment. It incorporates a representation of their home layout, sensor placement within the layout and the observed behaviour captured by the sensors (see Section~\ref{sub:ModelGenerator}).
    \item Properties: For each participant we define a set of properties which will be formally checked.  These properties are derived from existing safety guidelines and the 
    individual's contextual information, including needs and preferences, creating requirements for the expected behaviour. The properties are encoded in LTL and a model checker is used to verify that the model satisfies the specified requirements (see Section~\ref{sub:PropertySpecification}).
    \item NuSMV: This is the model-checking engine. It verifies the correctness of the formal model against specified properties and generates a sensor event trace (cf.\ a counterexample) highlighting the sequence of events leading to any violation.
\end{itemize} 

To streamline the pipeline, an automated tool is used to generate formal symbolic models based on home layouts and the positioning of sensors.

\subsection{Data Collection Process}
\label{sub:DataCollectionProcess}
 
Each potential participant is provided with an information sheet outlining the purpose and scope of the study.  Consent forms are used to ensure participants formally agree to taking part and that they understand their rights to withdraw at a later date if they so desire. Ethical approval for the study protocol was obtained from the School of Informatics Ethics Committee (application number 558534) at the University of Edinburgh prior to the start of recruitment.

In what follows, we describe the qualitative and sensor data collection. For different participants, data are collected over periods of one to three weeks. This duration was the result of a number of factors within the six-month period allocated for sensor deployment.  Installations were timed to avoid visitors, minimising noise caused by the presence of another person and ensuring data reflected typical ADLs. In addition, participants' plans (e.g.\ hospitalisation, relocation, holidays) further constrained the deployment period. Another factor was the availability of the team of researchers carrying out and managing the concurrent interviews and sensor deployments.

\subsubsection{Qualitative Data}
\label{subsec:QualitativeData}

\begin{figure}
    \centering
    \subfloat[]{\includegraphics[width=0.49\columnwidth]{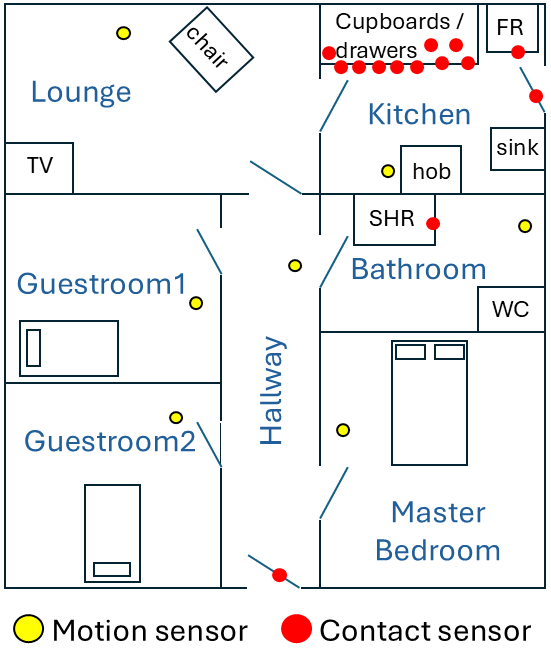}%
    \label{fig:ConcreteLayout}}
    \hfill
    \subfloat[]{\includegraphics[width=0.49\columnwidth]{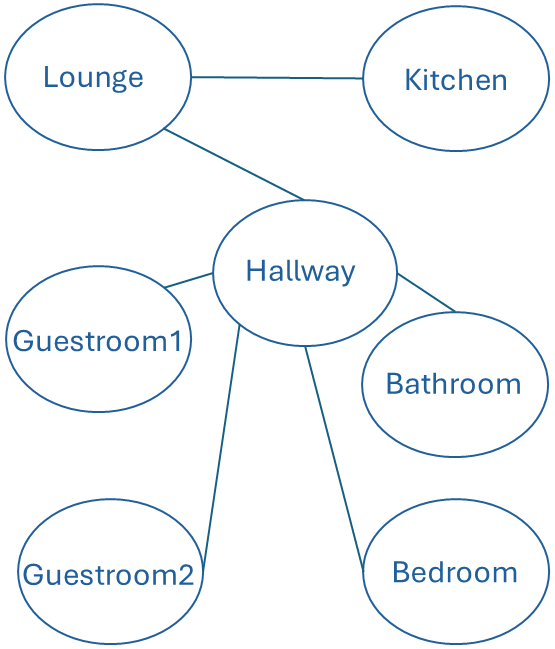}%
    \label{fig:AbstractRepresentation}}
    \caption{Concrete and abstract layouts of a participant's home. (a) Floorplan for a participant. (b) Graph representing the layout}
    \label{fig:Layout}
\end{figure}

Once consent is obtained, a home visit is scheduled with the participant for the collection of qualitative data. A semi-structured interview is used by the research team to gauge the participant's daily routines, the context in which these are performed and the needs and preferences the participant may have prior to sensors installation. The interview guide is divided into five sections: i) ADLs, ii) context and environment including the impact of sensors on the daily routine, iii) thoughts on privacy and security, iv) current needs and preferences, feedback and suggestions and v) general well-being. 
From the interview and observations, we also obtain insights into the possible placement of sensors. For example, some participants express discomfort with motion sensors in private areas such as bedrooms, but are comfortable with contact sensors in the same room. Figure~\ref{fig:ConcreteLayout} shows an example of the layout and the placement of sensors for one participant, based on their activities and privacy concerns. At the end of sensor data collection, a second interview is conducted to assess any changes in the participant’s routines or needs and to determine whether their behaviour was influenced by the presence of sensors.  Qualitative data are used to identify requirements that can be translated into properties regarding the participant's behaviour.

\subsubsection{Sensor Data}
\label{subsec:SensorData}

\begin{figure}[t]
    \centering    
    \subfloat[]{\includegraphics[width=0.45\columnwidth]{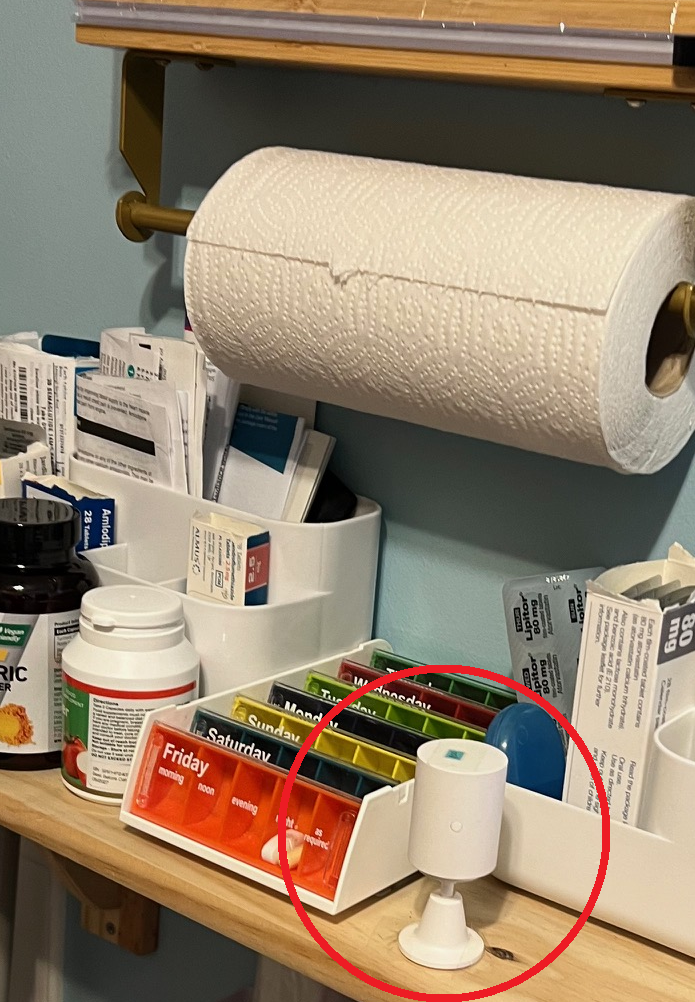}%
    \label{fig:MotionSensor}}
    \hfill
    \subfloat[]{\includegraphics[width=0.45\columnwidth]{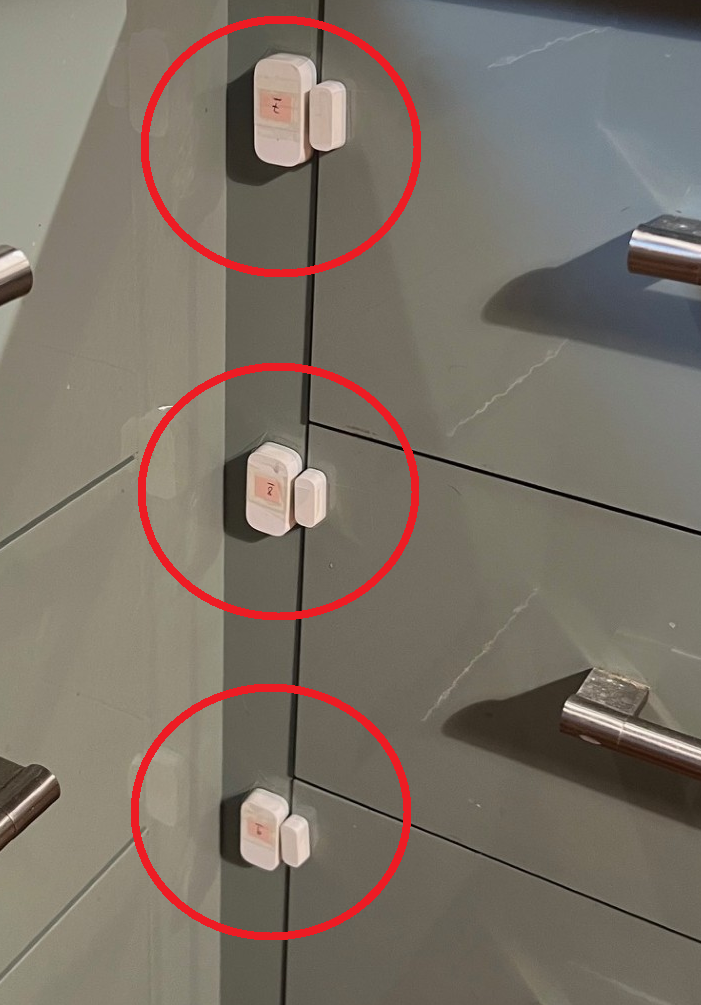}%
    \label{fig:ContactSensors}}
    \caption{Sensors installed in participants' home kitchens. (a)~Motion sensor medication. (b) Contact sensors on drawers.}
    \label{fig:SensorsPics}
\end{figure}

Sensor data are collected from contact and motion sensors placed throughout the participant’s home. The latter are positioned to cover living spaces and areas of transit, while contact sensors are attached to objects like cupboard doors or drawers.  Passive infrared (PIR) sensors are used to detect movement, with a detection angle of \ang{150} and a range of up to $7$ metres. These sensors have two possible states: \textit{clear} and \textit{detected}. If motion is registered while the sensor is in a \textit{clear} state, the state changes to \textit{detected} and remains in that state for $30$ seconds. If no further motion is registered within $30$ seconds, the sensor switches back to the \textit{clear} state. If motion is registered while the sensor is in the \textit{detected} state, the 30-second timer is restarted. Each contact sensor has two possible states: \textit{open} and \textit{closed}.  The contact sensor consists of two components: body and magnet.  When the body is within $8$~mm of the magnet, the state of the sensor is \textit{closed} and beyond this distance, the state is \textit{open}. Additionally, each contact sensor registers its internal temperature. Figure~\ref{fig:MotionSensor} shows a motion sensor used to detect movement in the area where medication is kept on a shelf in a participant’s kitchen. Figure~\ref{fig:ContactSensors}  shows contact sensors attached to various drawers in the participant’s kitchen. Both motion and contact sensors operate on an event-driven basis and are connected to a ConBee II Zigbee gateway. As previously mentioned, Home Assistant is used to orchestrate data acquisition from the sensors and the data are stored using InfluxDB. The automation platform, gateway and database run on an NVIDIA Jetson Nano with a quad-core ARM CPU, a $128$-core GPU, and $4$~GB of RAM, operating on a Linux-based system. The collected data are exported from the database in CSV format, resulting in dataset sizes ranging from $15$~MB to $35$~MB.

Data were collected from sensors deployed in $10$ homes, with $15$ to $24$ sensors per home and data collection periods of $7$ to $21$ days. There was no restriction on the number of sensors deployed in each home. Sensor placement and type were guided by qualitative analysis of participants’ routines and home layouts. Participants were older adults ($5$ women, $5$ men) aged $67$–$97$ years, recruited through community centres and library postings. The participants lived in areas ranging from the $5$\textsuperscript{th} to the $10$\textsuperscript{th} decile of the Scottish Index of Multiple Deprivation (SIMD) \cite{ScottishGovernment2020SIMD}. SIMD is a standard measure of relative deprivation across small geographic areas in Scotland, combining indicators such as income, health and crime. This indicates that all participants resided in areas from mid to least deprived. For this work, we present data from two participants from the 5\textsuperscript{th} and 7\textsuperscript{th} deciles, one male and one female, to demonstrate the approach rather than report participant-specific results. The evaluations are presented in Section \ref{sub:PropertiesofParticipants}.

\subsection{Practical Considerations in Sensor Deployment}

Data were collected using Zigbee sensors without relying on participants’ Wi-Fi, ensuring the setup was independent of home resources and did not interfere with personal connectivity. Overall, the gateway, Home Assistant and InfluxDB performed reliably. In three houses, a motion sensor malfunction was observed (i.e.\ one malfunction per house). The sensors went offline within $2$ to $24$ hours after installation. The data collected were retained, but the three motion sensors were excluded from the analyses. A similar minor challenge arose when one participant repositioned the body and magnet parts of two separate contact sensors too far apart, leaving these sensors permanently open. These experiences highlight the importance of planning for connectivity and on-site contingencies, and underscore the need for a degree of sensor redundancy (e.g.\ cross-sensor checking) to mitigate potential failures.
Additionally, our sensor data have inherent interpretational limitations; for instance, we cannot guarantee that events such as opening cutlery drawers, cabinets, or the refrigerator correspond directly to food preparation or consumption activities.

\section{Data Processing and Models}
\label{sec:DataProcessAndModels}

In this section we describe the data processing steps, the encoding of properties and the generation of the symbolic models. 

\subsection{Preprocessing}
\label{subsec:Preprocessing}
\subsubsection{Information from participants} 
Each participant is first allocated a unique identifier to ensure privacy and confidentiality. The mapping between participants’ personal details and their unique identifiers is kept separately under password protection and with controlled access.

The data generated from the interviews allow us to learn about each individual's ADLs, personal preferences and living context, which, in turn, help us model the properties that reflect general and specific ADLs. Framework Analysis \cite{Gale2013}, a qualitative analysis method, is applied to reveal patterns and themes in the interview data using NVivo software (Limivero, v14){~}\cite{nvivo2023}. Properties of ADLs are identified by considering both our questions during interviews 
and the routines participants highlighted as important. These properties are then categorised as {\it general} (common to everyone, like eating or washing), {\it individual-specific} (like pet care or hobbies), or {\it routine} but non-daily activities that impact health (like using medication). 
This approach allows us to have a detailed representation of the properties used to verify general and specific ADLs.

\label{subsub:Interviews}
\subsubsection{Sensors}
\label{subsub:Sensors}
Data from contact and motion sensors are first processed to reduce noise and retain events that reflect user interactions.

A contact sensor broadcasts its current state every time it changes from \textit{open} to \textit{closed}, \textit{closed} to \textit{open} or when there is a fluctuation of $1^\circ\text{C}$ in the device's temperature.  
Repetitions of the same state are taken to be noise in the data triggered by temperature fluctuations so we remove consecutive updates where the state does not change and consider only the first occurrence. This provides a more accurate representation of the participant's interaction. 

A motion sensor broadcasts its state whenever there is a transition between \textit{clear} and \textit{detected} states. If the state is \textit{detected} and no additional motion is registered within $30$ seconds, the sensor broadcasts an artificial \textit{clear} state after the $30$ seconds.  All \textit{clear} states are removed from the data as they do not reflect the actual cessation of motion. This provides a more accurate representation of movement, treating motion sensor events as instantaneous. In addition, if continuous \textit{clear} and \textit{detected} states occur for a motion sensor, with no changes in other sensors, only the initial \textit{detected} state of the motion sensor is considered and subsequent states are discarded.

\subsection{Model generator}
\label{sub:ModelGenerator}

NuSMV is used to build a model that represents the home and behaviour of the participant. The model is a transition system where changes in sensors can occur one at a time. This
is in line with what we observe in the data. The model consists of: i) states, which represent observed configurations of sensor states and where the participant is during the execution of the system; ii) transitions, that correspond to events, which describe how the system evolves from one state to another; and iii) the initial conditions that define the sensors' starting configuration. The model is verified against a set of logically-encoded properties (see Section~\ref{sub:PropertySpecification}).

When a property is violated, NuSMV generates a counter-example: a specific sequence of states that shows the conditions leading to the violation, making it easier to identify and understand its cause.

To produce a bespoke model for a participant we use our model generator, which takes as input a file that specifies the sensors installed and the layout of a participant’s home.  For each sensor, we specify its type, installation room, specific location and the name of the sensor in the model. Table~\ref{tab:SensorsLayout} presents a sample of the sensors deployed in the home previously depicted in Figure~\ref{fig:ConcreteLayout}. The input file also includes the abstract layout of the home that describes the spatial arrangement of the rooms. This corresponds to a directed graph, with each node representing a room and the edges representing the connections between rooms. Figure~\ref{fig:AbstractRepresentation} depicts the graph of the layout in Figure~\ref{fig:ConcreteLayout}.  

\begin{table}
\centering
\caption{Sample of sensors deployed in the home depicted in Figure~\ref{fig:ConcreteLayout}.}

\setlength{\tabcolsep}{5pt}
\begin{tabular}{|c|c|c|c|}
\hline
Type&
Room&
Location&
Name\\
\hline
Contact&
Kitchen&
Rubbish bin lid&
RubbishLid\\
\hline
Contact&
Kitchen&
Refrigerator door&
RefrigeratorDoor\\
\hline
Contact&
Kitchen&
Microwave door&
MicrowaveDoor\\
\hline
Motion&
Kitchen&
Facing cabinets&
MotionKitchen\\
\hline
Contact&
Kitchen&
Drawer cutlery&
DrawerCutlery\\
\hline
Contact&
Kitchen&
Cupboard food&
CupboardFood\\
\hline
Motion&
Hallway&
Facing hallway&
MotionHallway\\
\hline
Contact&
Bathroom&
Shower cubicle door&
ShowerDoor\\
\hline
Motion&
Bedroom&
Facing bed&
MotionBedroom\\
\hline
\end{tabular}
\label{tab:SensorsLayout}
\end{table}

\lstset{
    numbers=left,                 
    numberstyle=\tiny\color{gray}, 
    stepnumber=1,                 
    numbersep=1pt,                
    frame=none,                 
    basicstyle=\ttfamily\footnotesize, 
    captionpos=b,                 
    breaklines=true,              
    backgroundcolor=\color{white}, 
    keywordstyle=\color{blue},    
    commentstyle=\color{green},   
    stringstyle=\color{red},      
    xleftmargin=10pt,             
    xrightmargin=10pt,            
    showstringspaces=false,       
    literate = {&}{{$\land$}}1
}

\noindent
\begin{minipage}{\columnwidth}
\begin{center}
\small
\begin{lstlisting}[caption={Sensor contact module: state, activation and
deactivation.}, label={lst:sensorAspects}, captionpos=b, literate={¬}{{$\neg$}}1{\&}{{$\land$}}1]
ASSIGN
  init(state) := FALSE;
  init(activated) := FALSE;
  init(deactivated) := FALSE;
  next(activated) := ¬ state & next(state);
  next(deactivated) := state & ¬ next(state);
\end{lstlisting}
\end{center}
\end{minipage}

The behaviour of each sensor in the model is encapsulated in a module. An excerpt of the NuSMV code for the contact module is shown in Listing~\ref{lst:sensorAspects}. 
For each contact sensor three variables are defined: its state, activation and deactivation, which are initialised as \texttt{FALSE} (see lines $2$~-~$4$). In the module, the \texttt{state} of a contact sensor is a boolean variable where \texttt{FALSE} indicates the body and magnet of the sensor are nearby (e.g.\ drawers \textit{closed} as shown in Figure{~}\ref{fig:ContactSensors}) and \texttt{TRUE} indicates they are apart. The \texttt{activated} boolean variable is set to \texttt{TRUE} at the time step when the \texttt{state} transitions from \texttt{FALSE} ($\neg$ \texttt{state}) to \texttt{TRUE} (\texttt{state}), as shown in line $5$, and it remains \texttt{TRUE} for one time step only. Note that while \texttt{activated} represents interaction with the person and thus serves as evidence of their activities, the state of a sensor \textit{being} open -- which can last for more than one time step -- does not. The \texttt{deactivated} boolean variable is set to \texttt{TRUE} at the time step when the \texttt{state} transitions in the opposite direction, as shown in line $6$, also remaining \texttt{TRUE} for one time step only. Similarly, while \texttt{deactivated} represents interaction by a participant, the state of a sensor \textit{being} closed does not. 

The behaviour of a motion sensor in the model follows a similar structure but with two variables instead of three: its state and activation. The \texttt{active} boolean variable is set to \texttt{TRUE} at the time step when the \texttt{state} transitions from \texttt{FALSE} to \texttt{TRUE} and it remains \texttt{TRUE} for one time step only. However, unlike the contact sensor, the state is explicitly set to \texttt{FALSE} immediately after being \texttt{TRUE}.

The model is provided with a subtrace of events captured by the different sensors. This subtrace represents the observed behaviour of a participant during a specific period of time.  Consider a trace $\pi$ for a set of sensors $S$.
Given a pair of unique time markers, $a$ and $b$ -- manually inserted to mark the start and end of a time interval -- the subtrace is the set of all events, chronologically ordered, from $a$ to $b$, where $t(\cdot)$ denotes the time of an event: 
\begin{equation*}
    \text{subtrace}(\pi, a, b) = \left\{
    \begin{array}{@{}l @{\;\big|\;} l@{}}         
          e_k \in \pi        & t(a) \leq t(e_k) \leq t(b) \\
    \end{array}
    \right\}
\end{equation*}

\begin{figure}
    \centering    
    \includegraphics[width=\columnwidth]{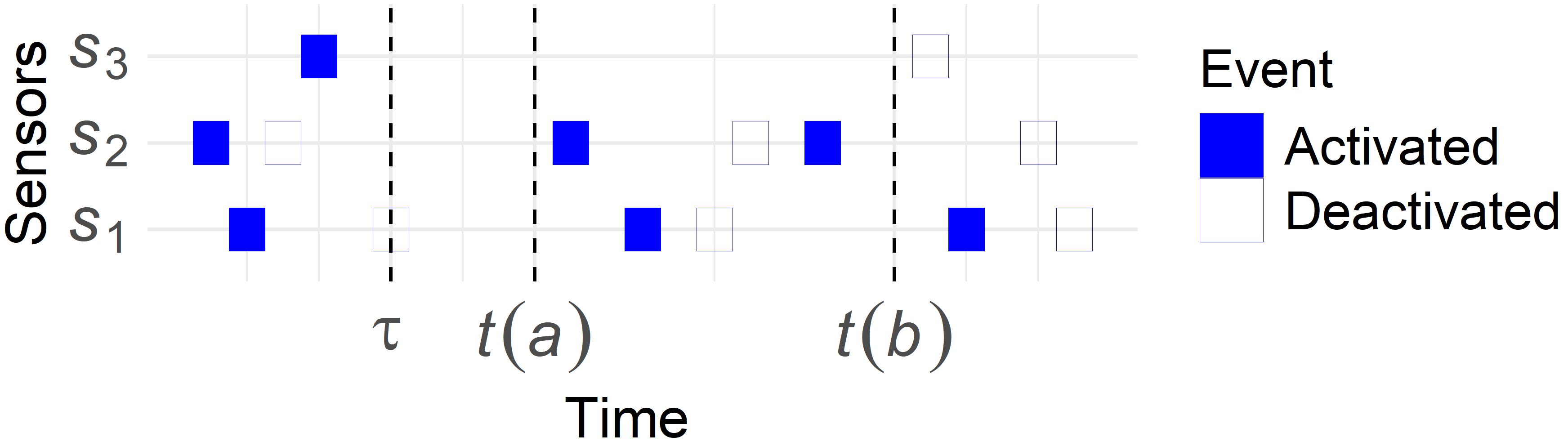}%
    \caption{Subtrace for the activation and deactivation of three sensors between time markers $a$ and $b$.}
    \label{fig:PlotSensors}
\end{figure}

In the model, an $n$-dimensional vector $V(\tau)$ represents the state of all $n$ sensors in a home at time $\tau$. For example, Figure{~}\ref{fig:PlotSensors} shows a subtrace in which a total of three sensors $S_1$, $S_2$ and $S_3$ change states over time. In this case, at time $\tau$, sensors $S_1$ and $S_2$ are deactivated while sensor $S_3$ is active in the corresponding 3-dimensional vector $V(\tau)$.

Using subtraces and the vector of states enables the analysis of the observed behaviour over specific intervals, helping mitigate the state space explosion problem, which refers to the rapid growth of possible system states in NuSMV. These intervals enable the identification of the specific times at which the behaviour was verified.

The model specifies the configuration of a group of sensors (such as in a room) by using a subvector $v(\tau)$ of $V(\tau)$. This allows us to capture a change in the room/group as a difference in the subvector in subsequent time steps. For example, grouping sensors including the shower cubicle door and the bathroom motion under the name Bathroom enables us to express a change in all these sensors as a single variable, $Bathroom.change$, in our model.  A detailed explanation of the various parts of a NuSMV model can be found in our previous work \cite{contreras2024}.

\subsection{Formalism and Property Specification}
\label{sub:PropertySpecification}

Our focus is on modelling behaviour over time. This makes the temporal logic LTL apt for our work as it allows us to specify properties using various temporal operators. In our work, we use LTL extended with
past-time operators~\cite{pastLTL}, henceforward referred to as past Linear Temporal Logic (pLTL), as it enables reasoning about events that have already occurred. This extension offers practical expressiveness, allowing properties based on sequencing of events to be specified more clearly and compactly.

We construct pLTL formulas, representing properties for a given participant (see next section), and then check whether all execution paths
(sequences of states) of the model satisfy them. A formula consists of atomic propositions (a property inherent in the state of the model), propositional operators and temporal operators.
Formally, formulas are generated according to the following BNF grammar:
\begin{equation*}
\label{log:LTL-formula}
\begin{aligned}
\phi, \psi ::= & \quad p \mid \neg \phi \mid \phi \lor \psi \mid \phi \land \psi \mid \phi \rightarrow \psi \\
              & \quad \mid X\phi
              \mid Y\phi\mid F\phi \mid O\phi \mid G\phi \mid \phi \mathrel{U} \psi \mid \phi \mathrel{R} \psi
\end{aligned}
\end{equation*}

Satisfaction of a formula $\phi$ by a path $\sigma$ at time index $i$ is denoted by $\sigma, i \vDash \phi$.
An atomic proposition is satisfied if the state at that index of the trace is labelled with it.
The non-temporal logical operators $\neg$, $\lor$, $\land$ and $\rightarrow$ have the same meaning as in propositional logic.

Some of the temporal operators of pLTL are described in Table~\ref{tab:LTL-temporal}. This selection does not preclude the use of other operators; rather, it reflects those that are specifically chosen for the properties discussed here, recognising that different properties may require different operators.   

\subsection{Behaviour-related Properties for Participants}
\label{sub:PropertiesofParticipants}

Properties are identified and categorised using the analysis of qualitative data mentioned
in Section{~}\ref{subsec:Preprocessing}. In what follows, we present three of these in case studies pertaining to two participants, referred to as participant A and participant B.  Each participant is in their late 60s, retired, living independently, not receiving external care and follows defined daily routines. This selection of properties does not preclude the inclusion of others; however, space constraints limit what we can present here.

\begin{table}
\caption{Sample of temporal operators used in pLTL}
    \centering
    \begin{tabular}{|p{56pt}|p{164pt}|}
    \hline
    Operator &  Interpretation\\
    \hline
    Next ($X$) & In $X\phi$, $\phi$ is expected to hold at the immediate next time step\\
    Previous ($Y$) & In $Y\phi$, $\phi$ held at the time step immediately before the current one\\
    Eventually ($F$) & In $F\phi$, $\phi$ will hold at some point in the future, though not necessarily immediately\\
    Once ($O$) & In $O\phi$, $\phi$ held at some point in the past, at or before the current time step\\
    Globally ($G$) & In $G\phi$, $\phi$ holds at the current time and continues to hold at all future time steps\\
    Until ($U$) & In $\phi \mathrel{U} \psi$, $\phi$ holds continuously up to a future point when $\psi$ becomes true. At that point $\psi$ must hold and $\phi$ must have held at every earlier step\\
    \hline
    \end{tabular}
    \label{tab:LTL-temporal}
\end{table}

A daily routine corresponds to a subtrace delimited by time markers $\mathit{startActivity}$ and $\mathit{endActivity}$. To improve clarity and reduce repetition in the specifications, the following pattern is defined:
\begin{equation}
\begin{gathered}
\mathit{WithinInterval}(\omega) \equiv \mathrel{~}{G} (\mathit{startActivity} \land\neg{~}\mathit{endActivity}\\
 \rightarrow \neg{~}\mathit{endActivity} \mathrel{U} (\omega \land\neg{~}\mathit{endActivity}))
\end{gathered}
\nonumber
\end{equation}
This expression specifies that, once the routine begins ($\mathit{startActivity}$), but before the routine ends ($\mathit{endActivity}$), the condition $\omega$ must eventually hold.

The following two properties describe Participant A’s early morning routine, which takes place in the home depicted in Figure{~}\ref{fig:Layout}.

\subsubsection*{Property 1} This specifies that participant A moves from the bedroom to the bathroom to take a shower. Since their preference is to do
this after waking up, we check this property in accordance to their morning behaviour
(see Section \ref{sec:ExperimentAnalysisResults}).  The property falls under the \textit{general} category of the qualitative analysis.  For the participant, performing this ADL involves moving from the bedroom to the hallway, then to the bathroom and closing the shower cubicle door. The property (\ref{prop:pLTL-example1}) specifies that motion in the bedroom ($\mathit{MotionBedroom.activated}$) holds until there is motion in the hallway ($\mathit{MotionHallway.activated}$), which in turn holds until there is a 
change in the bathroom ($\mathit{Bathroom.change}$). The change in the bathroom must persist until the shower cubicle door is closed ($\mathit{ShowerDoor.deactivated}$). The bedroom and the hallway each have a single motion sensor; in both cases, one sensor was sufficient to cover the entire area, given the layout. The bathroom has a motion sensor and a contact sensor. 
A change in the bathroom can be triggered by motion detected in the bathroom ($\mathit{MotionBathroom.activated}$) or by \textit{closing} ($\mathit{ShowerDoor.deactivated}$) or \textit{opening} ($\mathit{ShowerDoor.activated}$) the shower cubicle door.
\begin{equation}
\begin{gathered}
\mathit{WithinInterval}(\mathit{MotionBedroom.activated}\mathrel{U}\\(\mathit{MotionHallway.activated}\mathrel{U} (\mathit{Bathroom.change}{~}\\ \mathrel{U}\mathit{ShowerDoor.deactivated})))
\end{gathered}
\label{prop:pLTL-example1}
\end{equation}

Note that the property allows the participant to open the shower cubicle door (if closed), enter the cubicle, and then close the door to take a shower. This accounts for situations where the door may have been closed earlier, for example, after cleaning the shower.

\subsubsection*{Property 2}\label{lab:propertyPets} This specifies that participant A opens the refrigerator to get the salad to feed their pets (before interacting with any other cupboards). Note that since their preference is to do this before breakfast, we check this property in accordance to their morning behaviour (see Section{~}\ref{sec:ExperimentAnalysisResults}). From the qualitative analysis this is an {\it individual-specific} property. The property (\ref{prop:pLTL-example2}) specifies that if the refrigerator door is open ($\mathit{RefrigeratorDoor}.activated$) and the door has not been opened in any previous steps ($\mathrel{Y}(\neg\mathrel{O}\mathit{RefrigeratorDoor}.activated)$), then none of the drawers or cupboards have been previously opened or closed ($\neg \mathrel{O} \mathit{KitchenCupboards.change}$), where $\mathit{KitchenCupboards}$ represents  
all drawers and cupboards in the kitchen.
\begin{equation}
\begin{gathered}
\mathit{WithinInterval}(\mathit{RefrigeratorDoor.activated}{~}\land
\\ \mathrel{Y}(\neg \mathrel{O}\mathit{RefrigeratorDoor}.activated)\rightarrow\\ \neg \mathrel{O} \mathit{KitchenCupboards.change})
\end{gathered}
\label{prop:pLTL-example2}
\end{equation}

Note that this property allows the participant to move in the kitchen or to open or close the rubbish bin or the kitchen door, each having a contact sensor, without violating the property.

\subsubsection*{Property 3}\label{lab:propertyMedication} This corresponds to the \textit{routine} category of the qualitative analysis. It is formulated for participant B and describes the participant’s interaction with the medication dispenser located in the kitchen each morning within a defined time window specified by the participant, as detailed in the experiments (see Section{~}\ref{sec:ExperimentAnalysisResults}). This dispenser is monitored by a motion sensor, as shown in Figure{~}\ref{fig:MotionSensor}. To reach the kitchen, the participant must first pass through the lounge. The participant may move between the lounge and kitchen, and can interact with various sensor-equipped fittings before reaching the dispenser. The property (\ref{prop:pLTL-example3}) specifies that change in the lounge ($\mathit{Lounge_B.change}$) holds until there is change in the kitchen ($\mathit{Kitchen_B.change}$) and that change in the lounge or kitchen ($\mathit{Lounge_B.change} \mid \mathit{Kitchen_B.change}$) holds until the dispenser is accessed ($\mathit{MotionMedication_B.activated}$). 
\begin{equation}
\begin{gathered}
\mathit{WithinInterval}(\mathit{Lounge_B.change}\mathrel{U}  (\mathit{Kitchen_B.change}{~} \land \\ (\mathit{Lounge_B.change} \lor \mathit{Kitchen_B.change})\mathrel{U} \\\mathit{MotionMedication_B.activated}))
\end{gathered}
\label{prop:pLTL-example3}
\end{equation}

\subsubsection*{Additional Properties} Properties specifying the allowable sequence of events imposed by physical constraints were also encoded.  This was done to ensure the sensor events align with the physical layout, thereby verifying correct sensor functioning. We focus here on the physical layout for Participant A's home (see Figure{~}\ref{fig:Layout}).  Property (\ref{prop:physi-constraint1}) specifies that if motion is detected in the bedroom and, at a later time, in the bathroom, motion must also be detected in the hallway in between.

\begin{equation}
\begin{gathered}
\mathit{MotionBedroom.activated}{~}\land\mathrel{F} \mathit{MotionBathroom.activated} \rightarrow\\ \neg \mathit{MotionBathroom.activated}\mathrel{U}\mathit{MotionHallway.activated}
\end{gathered}
\label{prop:physi-constraint1}
\end{equation}

Property (\ref{prop:physi-constraint3}) specifies that when the participant moves from the lounge to the kitchen, the first detected event in the kitchen must be motion.
\begin{equation}
\begin{gathered}
\mathit{Kitchen.change}{~}\land \mathrel{Y} \mathit{MotionLounge.activated}{~}\land \\ \mathrel{Y(\neg O}\mathit{Kitchen.change})\rightarrow \\\mathit{MotionKitchen.activated}
\end{gathered}
\label{prop:physi-constraint3}
\end{equation}
\section{Experiments, Results and Analysis}\label{sec:ExperimentAnalysisResults}
In this section we present the results of the experiments conducted to verify whether the properties described in Section~\ref{sub:PropertySpecification} are satisfied for the observed behaviour of the participants.  Each property is individually evaluated on a model generated for the participant. The model includes the data captured from sensors (i.e.\ the subtrace) during the period of time identified as relevant for the property (e.g.\ the morning for Property \ref{prop:pLTL-example1}).

We use a sliding time window to segment the data into $20$ minute subtraces (see Section{~}\ref{sub:ModelGenerator}), which are then passed to the model, allowing for more precise identification of when a property holds or fails. To avoid cutting off a subtrace where a property might hold, we allow a $10$ minute overlap between subtraces. The time window was chosen pragmatically, as the identified ADLs were of short duration. Using shorter time windows also reduces the number of events per subtrace, helping to mitigate the known challenge of state space explosion problem and improve performance.

From the qualitative analysis, it is identified that participant A performs their morning activities around $8\text{:}30$AM every morning. To compensate for early and late mornings, we consider all data captured between $7\text{:}30$AM and $9\text{:}30$AM when evaluating the properties. Similarly, participant B performs their activities around $8\text{:}00$AM every morning; we consider all data captured between $7\text{:}00$AM and $9\text{:}00$AM.

Table{~}\ref{tab:ResultsParticipantA} summarises the outputs when verifying Property{~}\ref{prop:pLTL-example1} and Property{~}\ref{prop:pLTL-example2} for participant A. The table is organised with grouped columns for each property. The \textit{Status} column states whether the property was satisfied or violated within the morning routine. The \textit{Time} column indicates the time frame for which the property is satisfied. The \textit{Seq} column shows the number of subtraces identified within the interval frame for which the property is satisfied. 

Property{~}\ref{prop:pLTL-example1} was satisfied on all dates between $7\text{:}30$AM and $9\text{:}30$AM except August $28^{th}$. Analysis of the counterexample revealed that participant A followed the expected sequence of events between $7\text{:}50$AM and $8\text{:}20$AM, but once in the bathroom, returned to the hallway before going back to the bathroom and closing the shower cubicle door. This unexpected transition caused the property to be evaluated as \texttt{FALSE}, even though a morning shower activity was almost certainly completed. This indicates that the apparent violation arose from a minor deviation in the sequence, suggesting that temporal requirements may need refinement to accommodate minor deviations in daily routines.  

Similarly, Property{~}\ref{prop:pLTL-example2} was violated only on Sunday, $25^{th}$ August, when participant A did not open the refrigerator before $9\text{:}30$AM. On the same day, a slight delay in the shower routine was observed, indicating that weekend routines could be shifted to later times. This suggests that defining the start of the day relative to the participant’s first set of sensor activations is an aspect to be explored further. In addition a higher level of specificity may be needed when conducting interviews to address such nuances.

\begin{table}
\caption{Participant A}
\setlength{\tabcolsep}{5pt}
\begin{center}
\begin{tabular}{|c|c|c|c|c|c|c|}
\hline
\multirow{2}{*}{Date} & \multicolumn{3}{c|}{Property $1$ (morning shower)} & \multicolumn{3}{c|}{Property $2$ (refrigerator)} \\
\cline{2-7}
{~} & Status & Time & Seq & Status & Time & Seq\\
\hline
23-08-24 & Satisf. & 7:50--8:20 & 2 & Satisf. & 7:50--8:20 & 2\\
\hline
24-08-24 & Satisf. & 8:30--9:00 & 2 & Satisf. & 9:00--9:30 & 2\\
\hline
25-08-24 & Satisf. & 8:50--9:10 & 1 & Viol. & -- & 0\\
\hline
26-08-24 & Satisf. & 8:20--8:50 & 2 & Satisf. & 7:30--9:30 & 1\\
\hline
27-08-24 & Satisf. & 8:10--8:40 & 2 & Satisf. & 7:30--9:30 & 2\\
\hline
28-08-24 & Viol. & -- & 0 & Satisf. & 7:30--9:30 & 2\\
\hline
29-08-24 & Satisf. & 8:30--8:50 & 1 & Satisf. & 7:30--9:30 & 2\\
\hline
\end{tabular}
\end{center}
\label{tab:ResultsParticipantA}
\end{table}

The two physical properties specified for the layout of participant A were evaluated throughout the day from $7\text{:}00$AM to $8\text{:}00$PM. Property{~}\ref{prop:physi-constraint1} held throughout the day. Property{~}\ref{prop:physi-constraint3}, related to motion in the kitchen, was occasionally violated. The violations occurred when the participant entered the kitchen and opened the bin. We believe that these occasional violations occurred because motion near the bin was not properly captured by the motion sensor in the kitchen.

For participant B, we evaluated whether the scheduled medication intake (see Property{~}\ref{prop:pLTL-example3}) occurred while the participant moved between the kitchen and the lounge between $7\text{:}00$ and $9\text{:}00$. Table{~}\ref{tab:ResultsMedicine} shows the dates and time intervals for which the property was satisfied. The property was not satisfied on Friday November $22^{nd}$; participant B entered the lounge, a necessary step to reach the medication, only after $9\text{:}50$AM. While some ADLs allow for flexibility in when they occur, scheduled medication intake is a more rigid requirement. A deviation in this activity could indicate a health safety risk.

\begin{table}
\caption{Participant B (Property Medication)}
\setlength{\tabcolsep}{5pt}
\begin{center}
\begin{tabular}{|c|c|c|c|}
\hline
Date&
Status&
Time&
Seq\\
\hline
16-11-24&
Satisf.&
7:50 - 8:50&
4\\
\hline
17-11-24&
Satisf.&
8:20 - 8:50&
2\\
\hline
18-11-24&
Satisf.&
8:40 - 9:00&
1\\
\hline
19-11-24&
Satisf.&
7:20 - 8:20&
5\\
\hline
20-11-24&
Satisf.&
8:10 - 9:00&
4\\
\hline
21-11-24&
Satisf.&
7:00 - 7:40&
3\\
\hline
22-11-24&
Viol.&
--&
0\\
\hline
\end{tabular}
\end{center}
\label{tab:ResultsMedicine}
\end{table}

\subsubsection*{Limitations} This study has some limitations. Sensor placement was sometimes restricted by participant preferences, while pets and unexpected visitors occasionally introduced noise, as the models assume a single participant. The sensors provide only evidence of events (e.g.\ refrigerator opening) and do not guarantee inherent interpretation (e.g.\ food intake). Data collection was constrained by study duration, resources and participant availability; consequently, the study included a small number of participants. Malfunctions were observed in three motion sensors (about 1\% of the data), which were excluded from analysis. Importantly, though, our person-centred approach mitigates challenges, such as sensor placement, ensuring relevance to participants’ salient ADLs.

\section{Conclusion}
\label{sec:Conclusion}
In this paper, we have presented a framework for representing and reasoning about ADLs performed by OAs. As part of our work, we recruited $10$ participants aged $67$ to $97$. For each participant, we collected both qualitative and quantitative data over a period ranging from one to three weeks. Our approach leverages contextual information from the participants and data captured by commercially available sensors installed in their homes to build tailored models. We specify properties tailored to the needs and preferences expressed by the participants and encode them in LTL extended with past-time operators. Models are automatically generated for each participant and the model checker NuSMV is used to evaluate whether the observed behaviour of participants conforms to the encoded properties. 

To demonstrate the applicability of our framework, we conducted experiments on real-world datasets from two participant living independently. The results support the feasibility of our framework for analysing tailored properties and identifying deviations from expected behaviour. From the case studies presented, we observe that one deviation corresponds to changes in a participant's routine when taking a shower. Another deviation was triggered because the activity (medication intake) was not performed. Further analysis reveals that, although the activities are not performed at the expected time, they are carried out correctly outside the anticipated time period. Although this may suggest that the timing for property verification could be relaxed, timely performance of activities such as scheduled medication intake might be essential and a violation may indicate the need for an intervention. Overall, our initial findings show the potential of the framework to be used in the home of OAs living independently. However, even in specific settings where an individual’s context and their preferences and needs are taken into account, ADL monitoring remains challenging.  This is because human behaviour and sensing granularity can result in ambiguous interpretations, such as the act of opening a refrigerator not necessarily indicating food consumption.


\begin{thebibliography}{00}

\bibitem{Chen2023} C. Chen, S. Ding and J. Wang, ``Digital health for aging populations''. {\em Nature Medicine}. \textbf{29}, 1623-1630 (2023,7), \url{https://doi.org/10.1038/s41591-023-02391-8}

\bibitem{Tkatch2020ReducingLA} R. Tkatch, L. Wu, S. MacLeod, R. Ungar, L. Albright, D. Russell, J. Murphy, J. Schaeffer and C.S. Yeh, ``Reducing loneliness and improving well-being among older adults with animatronic pets''. {\em Aging \& Mental Health}. \textbf{25} pp. 1239 - 1245 (2020), \url{https://api.semanticscholar.org/CorpusID:218491804}

\bibitem{Lette2020} M. Lette, E. Ambugo, T. Hagen, G. Nijpels, C. Baan and S.R. de Bruin , ``Addressing safety risks in integrated care programs for older people living at home: a scoping review''. {\em BMC Geriatrics}. \textbf{20}, 81 (2020,2,28), \url{https://doi.org/10.1186/s12877-020-1482-7}

\bibitem{Khalili2024} G. Khalili, M. Zargoush, K. Huang and S. Ghazalbash, ``Exploring trajectories of functional decline and recovery among older adults: a data-driven approach''. {\em Scientific Reports}. \textbf{14}, 6340 (2024,3,15), \url{https://doi.org/10.1038/s41598-024-56606-0}

\bibitem{mmilnac2016} M. Mlinac and M. Feng. ``Assessment of Activities of Daily Living, Self-Care, and Independence''. {\em Archives Of Clinical Neuropsychology}. \textbf{31}, 506-516 (2016,8), \url{https://doi.org/10.1093/arclin/acw049}

\bibitem{clarke1999model} E. Clarke, O. Grumberg, D. Peled and D.A. Peled, Model Checking. (MIT Press,1999), \url{https://books.google.fr/books?id=Nmc4wEaLXFEC}

\bibitem{cimattiNuSMV} A. Cimatti, E. Clarke, E. Giunchiglia, F. Giunchiglia, M. Pistore, M. Roveri, R. Sebastiani and A. Tacchella, NuSMV 2: An OpenSource Tool for Symbolic Model Checking. {\em Computer Aided Verification}. pp. 359-364 (2002)

\bibitem{manna1992} Z. Manna and A. Pnueli. ``The temporal logic of reactive and concurrent systems''. (Springer-Verlag, 1992)

\bibitem{contreras2024} R. Contreras, F. Smola, J. Zheng, J. Hillston, J. Fleuriot ``Verifying properties of activities of daily living''. {\em Proceedings Of The 12th International Symposium DataMod 2024}. pp. 1-18 (2024,10).

\bibitem{datasetMunguia2004}
E. Munguia. \textit{Activity Recognition in the Home Setting Using Simple and Ubiquitous Sensors}. MIT, 2004.  
Available at: \url{https://courses.media.mit.edu/2004fall/mas622j/04.projects/home/}.

\bibitem{munguiatapia2003} E. Munguia, ``Activity recognition in the home setting using simple and ubiquitous sensors. (Massachusetts Institute of Technology, 2003)'', \url{http://courses.media.mit.edu/2004fall/mas622j/04.projects/home/Tapia03.pdf}

\bibitem{munguiatapia2004} E. Munguia, S. Intille and K. Larson, ``Activity Recognition in the Home Using Simple and Ubiquitous Sensors''. Lecture Notes in Computer Science. Springer, 2004, pp. 158–175

\bibitem{Rantz2013-ra} M. Rantz, M. Skubic, S. Miller, C. Galambos, G. Alexander, J. Keller, \& M. Popescu, Sensor technology to support Aging in Place. {\em J. Am. Med. Dir. Assoc.}. \textbf{14}, 386-391 (2013,6)

\bibitem{Kaye2011-ft} J. Kaye, S. Maxwell, N. Mattek, T. Hayes, H. Dodge, M. Pavel, H. Jimison, K. Wild, L. Boise, \& T. Zitzelberger, Intelligent Systems For Assessing Aging Changes: home-based, unobtrusive, and continuous assessment of aging. {\em J. Gerontol. B Psychol. Sci. Soc. Sci.}. \textbf{66 Suppl 1}, i180-90 (2011,7)

\bibitem{Kaye2017-gg} J. Kaye, Making pervasive computing technology pervasive for health \& wellness in aging. {\em Public Policy Aging Rep.}. \textbf{27}, 53-61 (2017,6)

\bibitem{Kaye2018-rd} J. Kaye, C. Reynolds, M. Bowman, N. Sharma, T. Riley, O. Golonka, J. Lee, C. Quinn, Z. Beattie, J. Austin, A. Seelye, K. Wild, \& N. Mattek, Methodology for establishing a community-wide life laboratory for capturing unobtrusive and continuous remote activity and health data. {\em J. Vis. Exp.}. (2018,7)

\bibitem{Skubic2015-di} M. Skubic, R. Guevara, \& M. Rantz, Automated health alerts using in-home sensor data for embedded health assessment. {\em IEEE J. Transl. Eng. Health Med.}. \textbf{3} pp. 2700111 (2015,4)

\bibitem{Janes2025-zw} W. Janes, N. Marchal, X. Song, M. Popescu, A. Mosa, J. Earwood, V. Jones, \& M. Skubic, Integrating ambient in-home sensor data and electronic health record data for the prediction of outcomes in amyotrophic lateral sclerosis: Protocol for an exploratory feasibility study. {\em JMIR Res. Protoc.}. \textbf{14} pp. e60437 (2025,3)

\bibitem{cooksMCI} G. Sprint, D.J. Cook, M. Schmitter-Edgecombe, and L.B. Holder. Multimodal fusion of smart home and text-based behavior markers for clinical assessment prediction. {\em ACM transactions on computing for healthcare} 3.4 (2022): 1-25.

\bibitem{kang2023} H. Kang, C. Lee, and S. Kang. ``A smart device for non-invasive ADL estimation through multi-environmental sensor fusion''. {\em Scientific Reports}. \textbf{13}, 17246 (2023,10,11), \url{https://doi.org/10.1038/s41598-023-44436-5}

\bibitem{matsui2020} T. Matsui, K. Onishi, S. Misaki, M. Fujimoto, H. Suwa and K. Yasumoto, ``SALON: Simplified Sensing System for Activity of Daily Living in Ordinary Home''. {\em MDPI Sensors}. \textbf{20} (2020), \url{https://www.mdpi.com/1424-8220/20/17/4895}

\bibitem{Tonkin2023} E. Tonkin,  . Holmes, H. Song, N. Twomey, T. Diethe, M. Kull, M. Perello, M. Camplani, S. Hannuna, X. Fafoutis, N. Zhu, P. Woznowski, G. Tourte, R. Santos, P. Flach \& I. Craddock. A multi-sensor dataset with annotated activities of daily living recorded in a residential setting. {\em Scientific Data}. \textbf{10}, 162 (2023), \url{https://doi.org/10.1038/s41597-023-02017-1}

\bibitem{konios2018} A. Konios, Y. Jing, M. Eastwood, and B. Tan, ``Unifying and Analysing Activities of Daily Living in Extra Care Homes''. {\em IEEE 16th Intl Conf On Dependable, Autonomic And Secure Computing(DASC)}. pp. 474-479 (2018)

\bibitem{garcia-Constantino2020} M. Garcia-Constantino, A. Konio, M.A. Mustafa, C. Nugent, G. Morrison, ``Ambient and Wearable Sensor Fusion for Abnormal Behaviour Detection in Activities of Daily Living''. {\em IEEE International Conference On Pervasive Computing And Communications}. pp. 1-6 (2020)

\bibitem{garciaPetri} M. Garcia-Constantino, A. Konios, and C. Nugent. ``Modelling Activities of Daily Living with Petri nets''. {\em IEEE International Conference On Pervasive Computing And Communications}. pp. 866-871 (2018)

\bibitem{magherini2013} T. Magherini, A. Fantechi, C.D. Nugent and E. Vicario, ``Using Temporal Logic and Model Checking in Automated Recognition of Human Activities for Ambient-Assisted Living''. {\em IEEE Transactions On Human-Machine Systems}. \textbf{43} pp. 509-521 (2013,11)

\bibitem{Kessler2025-gs} D. Kessler, S. Irons, M. Franz, N. Thomas, J. Kaye, M. Finlayson, \& F. Knoefel, Codesign of a framework to support compassionate care appropriate to technology use by people with cognitive decline. {\em Disabil. Rehabil.}. pp. 1-9 (2025,5)

\bibitem{Alexander2025-io} G. Alexander, A. Livingstone, S. Han, W. Chapman, T. Comans, G. Demiris, M. Fisk, M. Fossum, C. Fung, R. Kennedy, T. O'Malley, M. Skubic, \& IS-ITCOP Participants Emerging models of care using IT in long-term/post-acute care: A comparative analysis of human and AI-driven qualitative insights. {\em J. Gerontol. Nurs.}. \textbf{51}, 6-11 (2025,4)

\bibitem{guerra2023} B. Guerra, E. Torti, E. Marenzi, M. Schmid, S. Ramat, F. Leporati and G. Danese, ``Ambient assisted living for frail people through human activity recognition: state-of-the-art, challenges and future directions''. {\em Frontiers In Neuroscience}. \textbf{17} (2023,10)

\bibitem{robinson2020} E. Robinson, G. Park, K. Lane, M. Skubic, M. Rantz ``Technology for healthy independent living: Creating a tailored in-home sensor system for older adults and family caregivers'' {\em J. Gerontol. Nurs.}. \textbf{46}, 35-40 (2020,7)

\bibitem{zigbee2020} Connectivity Standards Alliance Zigbee Specification. Available at: \url{https://csa-iot.org/all-solutions/zigbee/}, Accessed: 2025-06-03.

\bibitem{homeassistant} Home Assistant Community. \emph{Home Assistant}, Version 2024.8. Available at: \url{https://www.home-assistant.io/}. Accessed 2025-02-06.

\bibitem{influxdb} InfluxData. \emph{InfluxDB: Time-Series Database}, Version 2.5.0. Available at: \url{https://www.influxdata.com/products/influxdb/}. Accessed 6 Feb 2025.

\bibitem{ScottishGovernment2020SIMD} {The Scottish Government. ``Scottish Index of Multiple Deprivation 2020}'', Available at: \url{https://www.gov.scot/collections/scottish-index-of-multiple-deprivation-2020}, Accessed: 2025-10-01.

\bibitem{Gale2013} N. Gale, G. Heath, E. Cameron, S. Rashid and S. Redwood ``Using the framework method for the analysis of qualitative data in multi-disciplinary health research''. \emph{BMC Medical Research Methodology}. \textbf{13}, 117 (2013,9,18), \url{https://doi.org/10.1186/1471-2288-13-117}

\bibitem{nvivo2023}Lumivero NVivo (Version 14) [Computer software].  (2023), Available from \url{https://www.lumivero.com/products/nvivo/}

\bibitem{pastLTL} M. Benedetti and A. Cimatti, ``Bounded Model Checking for Past LTL''. {\em Tools And Algorithms For The Construction And Analysis Of Systems}. pp. 18-33 (2003)

\end{thebibliography}
\end{document}